\documentstyle[12pt]{article}
\begin{document}
\vspace*{2.5 cm}
\begin{center}
{\bf \Large Inclusion of Diffraction Effects in the Gutzwiller Trace Formula}
\end{center}
\begin{center}
{\bf G. VATTAY$^a$, A. WIRZBA$^b$ and P.E. ROSENQVIST$^c$}
\end{center}
\begin{center}
{\rm \small $^a$ Institute for Solid State Physics, E\"otv\"os University,
M\'uzeum krt. 6-8, H-1088 Budapest, Hungary}

{\rm \small $^b$ Institut f\"ur Kernphysik, Technische Hochschule,
Schlo\ss gartenstr.\ 9, D-64289 Darmstadt, Germany}

{\rm \small $^c$ Niels Bohr Institute,
Blegdamsvej 17, DK-2100  Copenhagen \O, Denmark}
\end{center}

\vspace{15mm}

\baselineskip 7.5mm
\parbox[t]{160mm}{\baselineskip 7.5mm
{\bf Abstract} -
The Gutzwiller trace formula is extended to include
diffraction effects. The new trace formula involves
periodic rays which have non-geometrical segments as a result of
diffraction on the surfaces and edges of the scatter.}

\vspace{36pt}

Gutzwiller's trace formula$^1$ is an increasingly popular tool for
analyzing semiclassical behaviour. Recently, it has been demonstrated
that using proper mathematical apparatus, like Gutzwiller-Voros$^2$
zeta functions, cycle expansions$^3$ or quantum Fredholm determinants$^4$, the
trace formula can
successfully predict individual eigenenergies of bound systems and
resonances of open
scattering systems.
The physical content of the trace formula is the geometrical optical
approximation of quantum mechanics via canonical invariants of closed
classical orbits. This approximation
is very accurate when periodic orbits sufficiently cover the phase space of the
chaotic system.
This is not the case when the number of
obstacles is small or their distance is large compared to their
typical size. Such a problem
occurs where the wave length of a quantum mechanical
(or optical) wave is very large compared to the spatial variation of
a repulsive potential, e.g.\ at the boundaries of
microwave guides, optical fibers, superconducting squids, or circuits in the
ballistic  electron transport,
i.e.\ in most of the devices used for so-called macroscopic
quantum mechanical (or optical) experiments. In such cases it is important to
take into account
the next-to-geometrical effects. In Ref. 5 we have shown how the
{\em Geometric Theory of Diffraction} (GTD)
for hard core potentials can be incorporated in the periodic orbit theory.
(Since the space here is limited, we invite the reader to study reference 5
for more details and explicite formulas.)

The diffracted rays connecting two points in the configuration space
can be derived from an extension of  Fermat's variational principle
of classical mechanics$^6$.
The generalized principle requires new classes of curves:
We have to consider for each triplet
of integers $r,s,t\ge 0$ the class of curves ${\cal D}_{rst}$
with $r$ smooth arcs on the surface, $s$ points on the edges
and $t$
points on the vertices of the boundary or the discontinuity. The curves
of the GTD are those which make the classical action stationary within
one of the classes ${\cal D}_{rst}$. The class ${\cal D}_{000}$ corresponds
to the usual geometrical orbits.
Once we know the generalized ray connecting ${\cal A}$ and ${\cal B}$
we can compute semiclassically the
Green's function $G(q_{\cal A},q_{\cal B},E)$ by tracing the ray$^6$:

{\bf a,} On the geometrical segments of the ray, the Green's function
is given by the energy domain Van-Vleck propagator
\begin{equation}
G(q,q',E)=
\frac{2\pi}{(2\pi i \hbar)^{3/2}}D_{\mbox{V}}^{1/2}(q,q',E)
e^{\frac{i}{\hbar}S(q,q',E)-\frac{i}{2}\nu\pi},
\end{equation}
where $D_{\mbox{V}}(q,q',E)=|\det(-\partial^2 S/\partial q_i
\partial q_j')|/|\dot{q}||\dot{q}'|$ is the Van-Vleck determinant
and $\nu$ is the Maslov index.

{\bf b,} When the geometrical ray hits a surface, an edge  or
a vertex of the obstacle it creates a
source for the diffracted
wave. The strength of the source is proportional to
the Green's function at the incidence of the ray
\begin{equation}
Q_{\mbox{diff}}=DG_{\mbox{inci}}\ .\label{D}
\end{equation}
The diffraction constant $D$ depends on the local geometry and
the nature of the diffraction. It has been determined
in Ref. 6 from the asymptotic
semiclassical expansion of the exact solution in a
simple geometry$^{6,7}$. For
the surface diffraction (creeping) its form is
\begin{equation}
D_l= 2^{{1}/{3}} 3^{{-2}/{3}}\pi e^{5 i {\pi}/{12}}
\frac{(k \rho)^{{1}/{6}}}{Ai'(x_l) }\ .\label{DL}
\end{equation}
Here  $Ai'(x)$ is the derivative of the Airy function,
$k=\sqrt{2mE/\hbar}$
is the wave number, $\rho$ is the
radius of the obstacle at the source of the creeping ray and
$x_l$ are the zeroes of the Airy integral.
The index $l\geq 1$ refers to
the possibility of initiating creeping rays with different modes,
each with its own profile. In practice only the low modes
contribute to the Green's function.
For edge diffraction the diffraction constant is
\begin{equation}
D=\frac{\sin(\pi/n)}{n}
\left[\left(\cos(\pi/n)-\cos((\theta-\alpha)/n)\right)^{-1} -
\left(\cos(\pi/n)-\cos((\theta+\alpha+\pi)/n)\right)^{-1}\right],
\label{Edge}
\end{equation}
where $(2-n)\pi$ is the angle of the edge ($n$ is a real number),
$\alpha$ is the incident
angle and $\theta$ is the outgoing angle. For details we
refer to Ref. 6.
The source then initiates a ray propagating along the surface
(for creeping) or a ray starting at the edge of the obstacle
(edge diffraction).
When the creeping ray leaves the surface
its intensity can be calculated from the relation (\ref{D})
due to the
reversibility of the Green's function.
The total Green's function is then the {\em product} of the
Green's functions
and diffraction coefficients along the ray.
If for example we have geometrical propagation from
${\cal A}$ to ${\cal A'}$, then a surface creeping from
${\cal A'}$ to ${\cal B'}$ and then again a geometrical
propagation from ${\cal B'}$ to ${\cal B}$, the total
semiclassical Green's function is
\begin{equation}
G(q_{\cal A},q_{\cal B},E)= G(q_{\cal A},q_{\cal
A'},E)D_{{\cal A'}}G^{\mbox{creeping}}(q_{\cal A'},q_{\cal
B'},E)D_{{\cal B'}}  G(q_{\cal B'},q_{\cal B},E).
 \label{Gprod}
\end{equation}

Contrary to the pure geometrical case
the  semiclassical energy-domain Green's function for rays with diffraction
arcs
have a multiplicative composition law.
When we incorporate diffraction effects into the trace formula,
periodic rays with diffraction segments also contribute.
We can handle separately the pure geometric cycles and the cycles with
at least one diffraction arc or edge:
\begin{equation}
\mbox{Tr}\:  G(E) \approx \mbox{Tr}\:  G_{G}(E) + \mbox{Tr}\:  G_{D}(E),
\end{equation}
where $\mbox{Tr}\:  G_{G}(E)$ is the ordinary Gutzwiller trace formula,
while $\mbox{Tr} \:   G_{D}(E)$
is the new trace formula corresponding
to the non-trivial cycles
of the GTD.
$\mbox{Tr}\:  G_{D}(E)$ can be computed by using appropriate
Watson contour integrals$^7$. For technical details we
refer the reader to Refs. 5 and 8. If we denote by $q_i$, $i=1,\dots,n$
(with $q_{n+i}\equiv q_i$) the
points along
the closed cycle, where the ray  changes from diffraction to pure geometric
evolution or vice versa
the trace for cycles with
{\em at least one diffraction arc}
can be expressed as the product
\begin{equation}
\mbox{Tr}\:  G_{D}(E)=\frac{1}{i\hbar}\sum_{\mbox{Cycles}}T(E)\prod_{i=1}^{n}
                              {D}(q_i)G(q_i,q_{i+1},E),\label{tracef}
\end{equation}
where $T(E)$ is the time period of the primitive cycle
and $D(q_i)$
is the diffraction constant (\ref{DL}) at the point $q_i$.
$G(q_i,q_{i+1},E)$ is
alternatingly
the Van-Vleck propagator, if $q_i$ and $q_{i+1}$ are connected by pure
geometric
arcs, or is given by creeping propagator$^7$ in case $q_i$ and $q_{i+1}$ are
the boundary
points of  a creeping arc.
The cycles with diffractional parts have the special property
that their energy domain Green's functions
are {\em multiplicative} along the ray. This does not
hold for pure geometrical cycles.

The eigenenergies can be recovered from the Gutzwiller-Voros spectral
determinant$^2$ $\Delta(E)$, which is related to the trace formula as
\begin{equation}
\mbox{Tr}\:  G(E)=\frac{d}{dE}\ln \Delta(E).
\end{equation}
The full semiclassical determinant can be written as the
formal product of
two spectral determinants, one corresponding to pure geometrical cycles
and one to the new  diffraction cycles,
$\Delta(E)=\Delta_G(E)\Delta_D(E)$ due to the additivity of the
traces. The diffraction part of the spectral determinant is
\begin{equation}
\Delta_D(E)=\exp\left(-\sum_{p,r=1}^{\infty}\frac{1}{r}\prod_{i=1}^{n_p}
[ D(q^p_i)G(q^p_i,q^p_{i+1},E)]^r\right),
\end{equation}
where the summation goes over closed primitive
cycles $p$ and the repetition number $r$.

To demonstrate the importance of the diffraction effects
to the spectra, we have calculated the resonances of the scattering
system of two equally sized hard circular disks$^{5,8}$. This system has
a $C_{2v}$ symmetry with four one-dimensional irreducible
representations. In Fig. 1 we show the $B_1$ resonances.
In this system there is only
one geometrical unstable cycle along the line connecting
the centers of the disks.
In Fig.\ 1 we  see that the new formula
describes the leading resonances with a few percent
error, while the computation based on the geometrical cycle alone
would give completely false result.

In bound systems the edge diffraction is probably more important
than the creeping diffraction. Recently, in Ref. 9, an edge
diffraction cycle has been observed in the Fourier transformed
level density. The amplitude of such a diffraction term can
be calculated from (\ref{tracef}) using the diffraction constant
(\ref{Edge}).

{\small {\em Acknowledgement} - The authors are very grateful to
Predrag Cvitanovi\'c for encouragement.
P.E.R. thanks  SNF for support
and G.V. acknowledges the support of Phare
Accord H 9112-0378 and OTKA F4286 and 2090.}

\begin{center}
{\bf REFERENCES}
\end{center}
\baselineskip 5mm
\begin{enumerate}
\small
\baselineskip 5mm
\item M. C. Gutzwiller, J. Math. Phys. {\bf 12}, 343 (1971)
\item A. Voros, J. Phys. {\bf A21}, 685 (1988)
\item P. Cvitanovi\'c, B. Eckhardt, Phys.Rev.Lett.{\bf 63}, 823 (1989)
\item P. Cvitanovi\'c and G. Vattay, Phys. Rev. Lett. {\bf 71},
4138 (1993), P. Cvitanovi\'c, P. E. Rosenqvist, G. Vattay and H. H.
Rugh, CHAOS {\bf 3} (4), 619 (1993)
\item G. Vattay, A. Wirzba and P.E. Rosenqvist, {\em
Periodic Orbit Theory of Diffraction}, Preprint, Niels Bohr Institute
(1994 January)
\item J. B. Keller, J. Opt. Soc. Amer. {\bf 52} 116 (1962)
J. B. Keller, in {\em Calculus of Variations
and its Application} (American Mathematical Society, 1958) p.27
B. R. Levy and J. B. Keller, Comm. Pur. Appl. Math.
{\bf XII}, 159 (1959)
B. R. Levy and J. B. Keller, Cann. J. Phys. {\bf 38}, 128
(1960)
\item W. Franz, {\em Theorie der Beugung Elektromagnetischer
Wellen}, Springer Verlag, Berlin (1957); Z. Naturforschung {\bf 9a}, 705
(1954)
\item A. Wirzba, CHAOS {\bf 2}, 77 (1992);
Nucl. Phys. {\bf A560}, 136 (1993)
\item Y. Shimizu and A. Shudo, Preprint (1994)
\end{enumerate}

\begin{figure}
\caption[1]{\small $B_1$ Resonances of the two-disk system$^{5,8}$
with disk separation $R=6a$
in the complex $k$ plane in units of the inverse disk radius $a$.
The diamonds label the exact resonances, the crosses are their semiclassical
approximations from the new extended trace formula.
These resonances cannot be predicted from the Gutzwiller
trace formula.}
\end{figure}
\end{document}